\documentclass[10pt,conference]{IEEEtran}
\usepackage{amsfonts}
\usepackage{amssymb}
\usepackage{amsmath}

\usepackage{dsfont}
\usepackage{graphicx,subcaption}
\usepackage{bm}
\usepackage{cite}
\usepackage{bigstrut}
\usepackage{float}
\usepackage{balance}
\usepackage{lipsum}
\usepackage{psfrag}
\usepackage{xcolor}
\usepackage{xfrac}
\usepackage{hyperref}
\usepackage[ruled,vlined]{algorithm2e}
\usepackage{tikz}
\usepackage{pgfplots}
\pgfplotsset{compat=1.15}
\usepackage{mathrsfs}
\usepackage[justification=centering, font=footnotesize]{caption}
\usetikzlibrary{arrows}
\usetikzlibrary{decorations.pathreplacing}
\usetikzlibrary{fadings}
\usetikzlibrary{fit}

\newtheorem{lemma}{Lemma}

\newtheorem{remark}{Remark}

\usepackage{mathtools}
\usepackage{comment}
\SetKwInput{KwInput}{Input}                
\SetKwInput{KwOutput}{Output}

\begin{document}
\IEEEoverridecommandlockouts
\newcommand\norm[1]{\left\lVert#1\right\rVert}

\title{Secrecy Rate Maximization with Artificial Noise for Pinching-Antenna Systems}
\author{\IEEEauthorblockN{Pigi P. Papanikolaou\IEEEauthorrefmark{1}, Dimitrios Bozanis\IEEEauthorrefmark{1}, Sotiris A. Tegos\IEEEauthorrefmark{1},\\ Panagiotis D. Diamantoulakis\IEEEauthorrefmark{1}, George K. Karagiannidis\IEEEauthorrefmark{1}}

\IEEEauthorblockA{\IEEEauthorrefmark{1}Department of Electrical and Computer Engineering, Aristotle University of Thessaloniki, Greece\\}
\IEEEauthorblockA{e-mail: \{pigipapa, dimimpoz, tegosoti, padiaman, geokarag\}@auth.gr}
\vspace{-5mm}

}

\maketitle

\begin{abstract}

Security is emerging as a critical performance metric for next-generation wireless networks, but conventional multiple-input-multiple-output (MIMO) systems often suffer from severe path loss and are vulnerable to nearby eavesdroppers due to their fixed-antenna configurations. Pinching-antenna systems (PAS) offer a promising alternative, leveraging reconfigurable pinching antennas (PAs) positioned along low-loss dielectric waveguides to enhance channel conditions and dynamically mitigate security threats. In this paper, we propose an artificial noise (AN)-based beamforming scheme for downlink transmissions in PAS, with the goal of maximizing the secrecy rate. A closed-form solution is derived for the single-waveguide scenario, while an alternating optimization approach addresses more complex multiple waveguide setups. Numerical results show that the proposed scheme significantly outperforms conventional MIMO and existing PAS security schemes.

\end{abstract}

\begin{IEEEkeywords}
pinching-antennas, secrecy rate maximization, artificial noise, physical layer security 
\end{IEEEkeywords}

\section{Introduction}\label{sec:Intro}

Wireless communication systems have undergone significant advances in recent decades, driven by the relentless pursuit of higher data rates, improved reliability, and enhanced security. Among various technologies, multiple-input-multiple-output (MIMO) systems have played a crucial role by introducing spatial degrees of freedom (DoFs), enabling beamforming techniques, and significantly increasing spectral efficiency \cite{MIMO2}. However, conventional MIMO systems typically rely on fixed antenna configurations, which limit their flexibility to dynamically respond to changes in the propagation environment, such as obstacles, user mobility, or evolving network requirements.

Recently, the concept of dynamic wireless channel reconfiguration has emerged, leveraging technologies such as reconfigurable intelligent surfaces (RISs) \cite{RIS}, movable antennas \cite{MOVE}, and fluid antennas \cite{fluid}. Despite their advantages, these approaches still face limitations, including restricted reconfiguration capabilities, limited movement ranges, and challenges in dealing with severe large-scale path loss and line-of-sight (LoS) blockage.

In response to these challenges, pinching-antenna systems (PASs) are emerging as a promising flexible-antenna solution. Originally demonstrated by NTT DOCOMO \cite{DOCOMO}, PASs leverage low-loss dielectric waveguides and deploy small dielectric particles, i.e., pinching antennas (PAs), which can be flexibly activated and repositioned along the waveguide. This technology enables significant reconfiguration capabilities, including adjusting antenna positions, effectively mitigating path loss by establishing strong LoS links and improving spatial channel control with minimal complexity and cost \cite{10909665}. Recent research has further advanced PASs by addressing critical design and optimization challenges. In \cite{ding}, the ability of PASs to improve LoS links and dynamically mitigate large-scale path loss was demonstrated, while in \cite{arxigos}, efficient optimization techniques were developed to maximize downlink data rates by strategically adjusting PA positions. Complementing these efforts, \cite{wang} developed a comprehensive physical and signal modeling framework, incorporating joint transmit and pinching beamforming techniques to optimize system power efficiency. In addition, \cite{bereyhi} extended PASs to multi-user MIMO scenarios, proposing hybrid beamforming schemes that significantly increase the achievable sum-rate performance.

Although this fundamental research has provided important insights into PASs, it has focused primarily on maximizing data rates without considering the inherent security risks. Due to the broadcast nature of wireless communications, confidential transmissions are particularly vulnerable to eavesdropping, motivating the development of physical layer security (PLS) techniques. State-of-the-art (SotA) studies have begun to explore the integration of PLS into PAS. In \cite{huawai}, the authors proposed fractional programming and gradient-based algorithms to design the optimal beamforming matrices that maximize secrecy rates (SRs), while in \cite{brasil}, expressions for key security metrics such as secrecy failure probability and secrecy capacity were derived, highlighting the performance gains achievable through the strategic use of PAs.

In this work, motivated by the enhanced security capabilities achieved when artificial noise (AN) is combined with the dynamic reconfigurability of PAs, we propose an AN-based beamforming scheme to maximize the SR in PASs. Specifically, we introduce a secure downlink transmission framework in which both confidential signals and AN are simultaneously transmitted through one or multiple waveguides, each integrated with an individually reconfigurable PA. Unlike conventional antenna systems (CASs), our approach exploits the spatial DoF provided by the PAS to jointly optimize beamforming vectors, AN covariance matrices, and PA positions, with the goal of strengthening the legitimate channel while effectively suppressing information leakage to potential eavesdroppers.
A closed-form solution is derived for the single-waveguide scenario, while the more complex multi-waveguide problem is addressed using an alternating optimization approach. Numerical results demonstrate the superiority of the proposed scheme over CAS and other SotA PAS security schemes.


\section{System Model and Problem Formulation}\label{sec:SysMod}

\begin{figure}
    \centering
    \includegraphics[width=1\columnwidth]{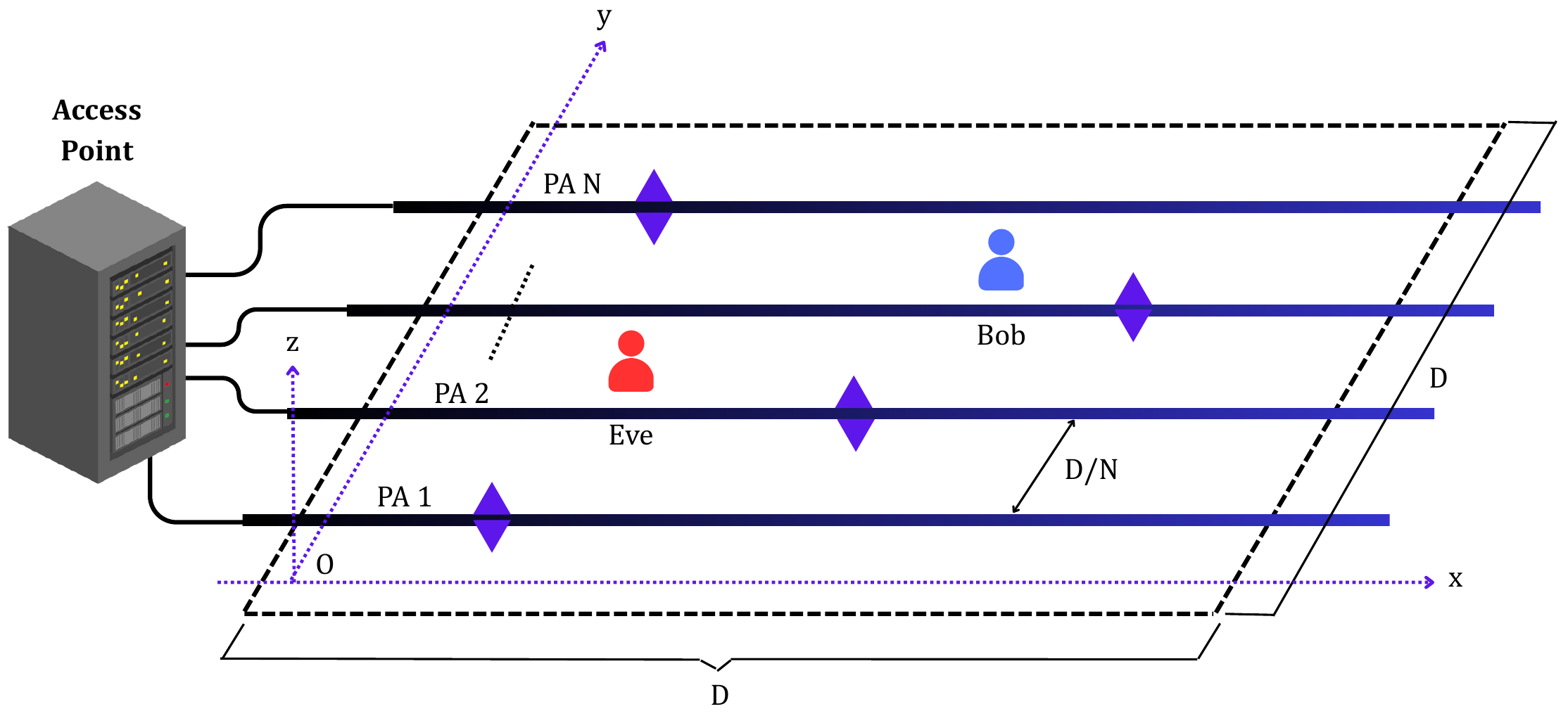}
    \vspace{-2mm}
    \caption{Illustration of a downlink PAS with multiple waveguides.}
    \vspace{-4mm}
    \label{fig:sys_model}
\end{figure}

In this work, we consider a secure downlink transmission framework based on PAs, as illustrated in Fig. \ref{fig:sys_model}. The base station (BS) is equipped with $N$ RF chains, each connected to a single waveguide via a flexible cable. On each waveguide, a single PA is activated by deploying a small dielectric particle whose position can be dynamically adjusted along the waveguide. The assumption of a single PA per waveguide, which is common in the literature \cite{wang,bereyhi,huawai}, is adopted because multiple PAs on a waveguide share the same signal, so they are easily coupled. This configuration was chosen to allow for independent phase tuning and full potential for spatial reconfiguration. At the BS, the confidential information is first processed via baseband beamforming, where it is superimposed with artificial noise to enhance security, and then routed to the RF chains for upconversion and transmission. This reconfigurability enables the BS to fine-tune the spatial transmission pattern, thereby enhancing secrecy by strengthening the legitimate link and hindering eavesdropping attempts. Without loss of generality, all waveguides are assumed to be aligned parallel to the x-axis, uniformly spaced, and mounted at a fixed height $d$. In our scenario, both the legitimate receiver (Bob) and the eavesdropper (Eve) are equipped with a single antenna and are distributed within a square region on the $x-y$ plane with a side length of $D$, with positions $\boldsymbol{\psi}_{B} = [x_b, y_b, 0]$ and $\boldsymbol{\psi}_{E} = [x_e, y_e, 0]$, respectively. Moreover, the length of each waveguide is assumed to be $D$ to ensure complete coverage of the area. The $y$-axis coordinate of the $n$-th waveguide is given as $\tilde{y}_{0,n}=(n-1)D/N$, thus the position of the $n$-th PA can be expressed as $\boldsymbol{\tilde{\psi}}_{P,n}=[\tilde{x}_{P,n},\:\tilde{y}_{P,n},\:d]$, where $\tilde{y}_{0,n}=\tilde{y}_{P,n}$.

\subsection{System Model}

The legitimate channels between all PAs and Bob are denoted as $\boldsymbol{h}_{B}=[h_{b,1}, ..., h_{b,N}]^T$, where $h_{b,n}$ represents the channels between the PA of the $n$-th waveguide with Bob, and it can be modeled as
\begin{equation}\label{channel}
    h_{b,n} = \frac{\eta^{\frac{1}{2}}e^{-j\frac{2\pi}{\lambda}\lVert \boldsymbol{\psi}_{B}-\tilde{\boldsymbol{\psi}}_{P,n} \rVert - j\frac{2\pi}{\lambda_g}\lVert \tilde{\boldsymbol{\psi}}_{0,n}-\tilde{\boldsymbol{\psi}}_{P,n} \rVert}}{\lVert \boldsymbol{\psi}_{B}-\tilde{\boldsymbol{\psi}}_{P,n} \rVert},   
\end{equation}
where $\tilde{\boldsymbol{\psi}}_{0,n} = [\tilde{x}_{0,n}, \tilde{y}_{0,n}, d]$ denotes the location of the feed point of the $n$-th waveguide, and $\lambda_g = \frac{\lambda}{n_{\mathrm{eff}}}$ denotes the guided wavelength with $n_{\mathrm{eff}}$ being the effective refractive index of a dielectric waveguide. As observed from \eqref{channel}, the channel in the PAS is affected by free-space path loss and dual phase shifts caused by signal propagation in free space and within the waveguide.

The transmitted signal $\boldsymbol{x}$ can be written as
\begin{equation}
    \boldsymbol{x} = \boldsymbol{w}s + \boldsymbol{m},
\end{equation}
where $\boldsymbol{w}\in \mathbb{C}^{N \times 1}$ is the beamforming vector, $s$ is the unit-power communication symbol, $\boldsymbol{m}\in \mathbb{C}^{N \times 1}$ is the AN matrix generated by the transmitter to interfere potential eavesdroppers. We assume that   $\boldsymbol{m}~\sim \mathcal{CN}(\boldsymbol{0}, \boldsymbol{R}_m)$, where $\boldsymbol{R}_m \succeq \boldsymbol{0}$ denotes the covariance matrix of the AN. Thus, the received signal at Bob can be expressed as 
\begin{equation}\label{receive}
    y_B = \boldsymbol{h}_B^H\boldsymbol{x}+z_B,
\end{equation}
where $z_B$ denotes the additive white Gaussian noise (AWGN) with variance $\sigma_B^2$.

The wiretap channels from all PAs to Eve are denoted as $\boldsymbol{h}_{E}=[h_{e,1}, ..., h_{e,N}]^T$, where $h_{e,n}$ represents the channels between the PA of the $n$-th waveguide with Eve. It should be highlighted that \eqref{channel} and \eqref{receive} also apply to Eve, whose position is given by $\boldsymbol{\psi}_{E}$, channel vector by $\boldsymbol{h}_{E}$, and AWGN variance is $\sigma_E^2$.

Therefore, the data rates for Bob and Eve are given as
\begin{equation}
\begin{split}
   R_\mathrm{Bob} =&  \log_2\left(1+\frac{\lVert \boldsymbol{h}_{B}^H\,\boldsymbol{w}\rVert^2}
    {\lVert \boldsymbol{h}_{B}^H\,\boldsymbol{R}_m\,\boldsymbol{h}_{B}\rVert + \sigma_B^2}\right),\\
   R_\mathrm{Eve} =&  \log_2\left(1+\frac{\lVert \boldsymbol{h}_{E}^H\,\boldsymbol{w}\rVert^2}
    {\lVert \boldsymbol{h}_{E}^H\,\boldsymbol{R}_m\,\boldsymbol{h}_{E}\rVert + \sigma_E^2}\right),
\end{split}
\end{equation}
which can be equivalently written as
\begin{equation}
\begin{split}\label{eq:R_alt}
   R_\mathrm{Bob} =&  \log_2\left(1+\frac{\operatorname{Tr}\left(\boldsymbol{H}_{B}\,\boldsymbol{W}\right)}
    {\operatorname{Tr}\left(\boldsymbol{H}_{B}\,\boldsymbol{R}_m\right) + \sigma_B^2}\right),\\
   R_\mathrm{Eve} =&  \log_2\left(1+\frac{\operatorname{Tr}\left(\boldsymbol{H}_{E}\,\boldsymbol{W}\right)}
    {\operatorname{Tr}\left(\boldsymbol{H}_{E}\,\boldsymbol{R}_m\right) + \sigma_E^2}\right),
\end{split}
\end{equation}
where $\operatorname{Tr}(\cdot)$ denotes the trace operator, while $\boldsymbol{H}_{B} = \boldsymbol{h}_{B} \boldsymbol{h}_{B}^H$, $\boldsymbol{H}_{E} = \boldsymbol{h}_{E} \boldsymbol{h}_{E}^H$ and $\boldsymbol{W}=\boldsymbol{w} \boldsymbol{w}^H$, thus the SR is given by
\begin{equation}\label{eq:SR}
    \mathrm{SR} = [R_\mathrm{Bob}-R_\mathrm{Eve}]^+.
\end{equation}

\subsection{Problem Formulation}

Assume that eavesdroppers pretend to be legitimate users and transmit uplink pilot signals. Consequently, the BS can obtain the channel state information (CSI) for both Bob and Eve \cite{huawai, optimization2}. By exploiting this prior knowledge, the BS can enhance the signal quality for Bob while effectively mitigating the information leakage to Eve. In this paper, we assume that perfect CSI is available.
Furthermore, as indicated by the SR expression, the additional spatial DoFs provided by the PAs allow not only the design of the baseband beamformer and the AN matrix, but also the reconfiguration of the channels. This flexibility allows the joint optimization of $\boldsymbol{W},\boldsymbol{R}_m$ and the PA position vector $\boldsymbol{\tilde{x}}_P = [\tilde{x}_{P,1}, \ldots, \tilde{x}_{P,N}],$ which contains the positions of all $N$ PAs. Thus, from \eqref{eq:R_alt}, \eqref{eq:SR}, and by defining the PA index set as $\mathcal{N} = \{1, 2, \ldots, N\},$ the optimization problem can be formulated as 
\begin{equation*} \tag{\textbf{P1}}\label{eq:basic_opt}
    \begin{array}{cl}
    \mathop{\max}\limits_{\boldsymbol{W}, \boldsymbol{R}_m, \boldsymbol{\tilde{x}}_P} &\mathrm{SR} \\
        \textbf{s.t.} & \mathrm{C}_1: \, \tilde{x}_{P,n}\in [0, D],\quad \forall\, n\in\mathcal{N},  \\
        & \mathrm{C}_2: \operatorname{Tr}\Bigl(\boldsymbol{W} + \boldsymbol{R}_m\Bigr) \leq P, \\
        & \mathrm{C}_3: \boldsymbol{W}\succeq 0,\quad \boldsymbol{R}_m\succeq 0,
    \end{array}
\end{equation*}
where $P$ denotes the total transmit power available at the BS. Constraint $\mathrm{C}_1$ ensures that the optimized location of the PA remains within the physical boundaries of the room, while constraint $\mathrm{C}_2$ guarantees that the total transmit power exactly meets the allocated power budget. Finally, constraint $\mathrm{C}_3$ requires that both $\boldsymbol{W}$ and $\boldsymbol{R}_m$ are positive semidefinite.

\section{AN-aided Secure Beamforming with a Single Waveguide}
In this section, we examine a special case of problem~\eqref{eq:basic_opt} in which the system consists of a single waveguide, i.e., \(N = 1\). In this scenario, both \(\boldsymbol{w}\), \(\boldsymbol{R}_m\) and $\boldsymbol{\tilde{x}}_P$ reduce to scalars, yielding a more tractable formulation. The resulting optimization problem can be expressed as
\begin{equation*} \tag{\textbf{P2}}\label{eq:opt_single}
    \begin{array}{cl}
    \mathop{\max}\limits_{w, R_m, \tilde{x}_P} &\mathrm{SR} \\
        \text{\textbf{s.t.}}& \mathrm{C}_1: \, \tilde{x}_P\in [0, D], \\
        & \mathrm{C}_2: \lVert w \rVert^2+R_m \leq P. \\
    \end{array}
\end{equation*}
We note that the objective in~\eqref{eq:opt_single} remains non-convex. To address this challenge, we propose an alternating optimization method that decomposes the original problem into two convex subproblems which are solved iteratively until convergence. They are formulated as 
\begin{equation*} \tag{\textbf{P2.1}}\label{eq:opt_single_x}
    \begin{array}{cl}
    \mathop{\max}\limits_{\tilde{x}_P} &\mathrm{SR}  \\
        \text{\textbf{s.t.}}& \mathrm{C}_1: \, \tilde{x}_P\in [0, D], \\
    \end{array}
\end{equation*}
where $w$ and $R_m$ are considered fixed, and
\begin{equation*} \tag{\textbf{P2.2}}\label{eq:opt_single_w}
    \begin{array}{cl}
    \mathop{\max}\limits_{w, R_m} &\mathrm{SR} \\
        \text{\textbf{s.t.}}& \mathrm{C}_2: \, \lVert w \rVert^2+R_m \leq P, \\
    \end{array}
\end{equation*}
where $\tilde{x}_P$ is now fixed.

\begin{lemma}\label{l1}
The objective function in optimization problem~\eqref{eq:opt_single_x} is univariate, and a closed-form solution for $\tilde{x}_P$ is derived for given values of $w$ and $R_m$.
\end{lemma}
\begin{IEEEproof}
The proof is presented in Appendix A.
\end{IEEEproof}

Having determined the optimal $\tilde{x}_P$ from~\eqref{eq:opt_single_x}, we now address problem~\eqref{eq:opt_single_w}. Specifically, we first solve constraint $\mathrm{C}_2$ for $R_m$ and by substituting this expression into equation~\eqref{eq:SR} and performing some algebraic manipulations, the objective function can be written as
\begin{equation}\label{eq:SR_fixed_x}
    \mathrm{SR} = \log_2{\left( \frac{(\eta P+r_B^2 \sigma_B^2)(\eta R_m+r_E^2 \sigma_E^2)}{(\eta P+r_E^2 \sigma_E^2)(\eta R_m+r_B^2 \sigma_B^2)}  \right)},
\end{equation}
where $r_B = \lVert \boldsymbol{\psi}_{B}-\tilde{\boldsymbol{\psi}}_{P} \rVert$ and $r_E = \lVert \boldsymbol{\psi}_{E}-\tilde{\boldsymbol{\psi}}_{P} \rVert$. 
\begin{lemma}\label{l2}
The objective function defined in \eqref{eq:SR_fixed_x} is monotone with respect to $R_m$. Specifically, if $r_B^2 \sigma_B^2 < r_E^2 \sigma_E^2$, the objective function is monotonically decreasing, and thus its maximum occurs at $R_m = R_m^{\mathrm{min}} = 0$. Conversely, if $r_B^2 \sigma_B^2 > r_E^2 \sigma_E^2$, it is monotonically increasing and has its maximum at $R_m = R_m^{\mathrm{max}} = P$.
\end{lemma} 
\begin{IEEEproof} 
The proof is given in Appendix B.
\end{IEEEproof} 
\begin{remark} 
In the proof of Lemma \ref{l1} for the case where the AWGN variances are equal (i.e., $\sigma_B^2 = \sigma_E^2$), a common assumption in the literature, it follows that the optimal strategy is to allocate the entire power budget to Bob without injecting any artificial noise, provided that Bob is physically closer to the transmitting antenna than Eve, since the condition simplifies to $r_B<r_E$. This condition is generally achievable through the reconfigurability of PAs. However, when operating with a single waveguide, it is not always feasible to ensure a PA positioning that places Bob closer to the transmit source than Eve. Consequently, the use of additional waveguides becomes imperative to ensure the desired proximity advantage for secure communications.
\end{remark}

\section{AN-aided Secure Beamforming with Multiple Waveguides}
Observing \eqref{eq:basic_opt}, we identify two major challenges. First, the objective function is inherently non-convex, making it difficult to find a global optimum. Second, the tight coupling of the optimization variables further increases the complexity of the problem. In this section, we introduce an alternating optimization scheme designed to overcome these challenges and effectively address problem~\eqref{eq:basic_opt}.

The goal is to maximize $\mathrm{SR}$ by jointly optimizing $\boldsymbol{W}$, $\boldsymbol{R}_m$, and the placement of the PAs along each waveguide. Notably, the non-convexity in problem~\eqref{eq:basic_opt} arises solely from the objective function. To address this challenge, we adopt an alternating optimization approach that decomposes the original problem into two more tractable subproblems, which are formulated as 
\begin{equation*} \tag{\textbf{P1.1}}\label{eq:basic_opt_fixed_x}
    \begin{array}{cl}
    \mathop{\max}\limits_{\boldsymbol{W}, \boldsymbol{R}_m} &\mathrm{SR} \\
        \text{\textbf{s.t.}}& \mathrm{C}_1: \, \operatorname{Tr}( \boldsymbol{W} + \boldsymbol{R}_m) = P, \\
        & \mathrm{C}_2: \boldsymbol{W}\succeq 0, \boldsymbol{R}_m\succeq 0,
    \end{array}
\end{equation*}
where the positions of the PAs, $\boldsymbol{\tilde{x}}_P$, are held fixed, and
\begin{equation*} \tag{\textbf{P1.2}}\label{eq:basic_opt_fixed_w}
    \begin{array}{cl}
    \mathop{\max}\limits_{\boldsymbol{\tilde{x}}_P} &\mathrm{SR} \\
        \text{\textbf{s.t.}}& \mathrm{C}_1: \, \tilde{x}_{P, n}\in [0, D], \forall n\in \mathcal{N},  \\
    \end{array}
\end{equation*}
where $\boldsymbol{W}$ and $\boldsymbol{R}_m$ are fixed, and the optimization performed only over the PA positions. These subproblems are then solved iteratively until convergence to a stable solution.

However, the objective function of the problem~\eqref{eq:basic_opt_fixed_x} is still non-convex. To this end, we introduce a slack variable $\gamma$, such that $R_{\mathrm{Eve}}\leq \log_2 \gamma$, thus \eqref{eq:basic_opt_fixed_x} can be reformulated as
\begin{equation*} \tag{\textbf{P1.1.1}}\label{eq:optimization_problem_multi_1_1}
    \begin{array}{cl}
    \mathop{\max}\limits_{\gamma, \boldsymbol{W}, \boldsymbol{R}_m}&\log_2\left(\frac{\operatorname{Tr}(\boldsymbol{H}_B\boldsymbol{R}_m)+\operatorname{Tr}(\boldsymbol{H}_B\boldsymbol{W})+\sigma_B^2}{\gamma (\operatorname{Tr}(\boldsymbol{H}_B \boldsymbol{R}_m)+\sigma_B^2)}\right) \\
        \text{\textbf{s.t.}}& \mathrm{C}_1: \, \operatorname{Tr}( \boldsymbol{W} + \boldsymbol{R}_m) = P, \\
        & \mathrm{C}_2: \boldsymbol{W}\succeq 0, \boldsymbol{R}_m\succeq 0, \\
        & \mathrm{C}_3: \operatorname{Tr}(\boldsymbol{H}_E \boldsymbol{W})\leq (\gamma -1)\left[\operatorname{Tr}(\boldsymbol{H}_E \boldsymbol{R}_m)+\sigma_E^2\right],\\
        & \mathrm{C}_4: \gamma \geq1.
    \end{array}
\end{equation*}


To solve problem~\eqref{eq:optimization_problem_multi_1_1}, we propose an alternating optimization framework that decouples the optimization of $\gamma$ from that of $\boldsymbol{W}$ and $\boldsymbol{R}_m$. In the first step, we optimize $\gamma$ for fixed $\boldsymbol{W}$ and $\boldsymbol{R}_m$. This subproblem is concave, which allows us to efficiently obtain its global optimum using numerical methods. In the second step, with $\gamma$ fixed, we optimize $\boldsymbol{W}$ and $\boldsymbol{R}_m$. This subproblem corresponds to a conventional SR maximization problem for which efficient solution techniques have been established in \cite{optimization2} by transforming the problem into an equivalent convex form, thus ensuring convergence to the globally optimal solution. Additionally, to ensure that the resulting beamforming matrix $\boldsymbol{W}$ is rank-1, an eigenvalue decomposition is applied as a refinement step.

Turning our attention to problem \eqref{eq:basic_opt_fixed_w}, we note that its inherently non-convex structure presents substantial challenges for standard optimization methods. To overcome these difficulties, we employ an exhaustive grid-search strategy, similar to that employed in \cite{wang,bereyhi,huawai} for similar PAS problems. This systematic exploration of the solution space allows us to accurately identify the global maximum of the objective function by optimally configuring the PA positions.

\section{Numerical Results}\label{sec:Num}

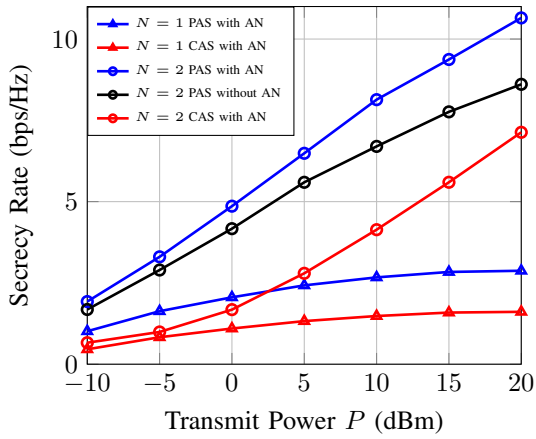
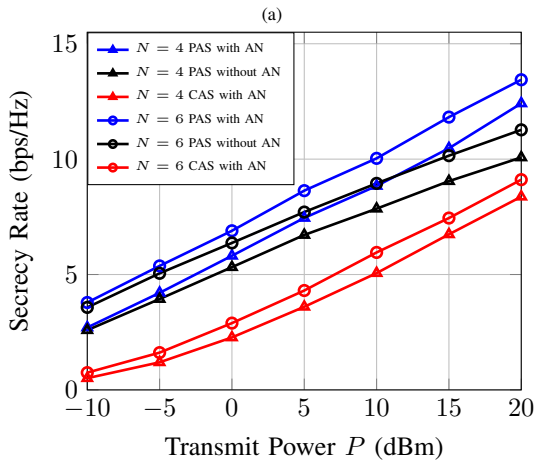
\begin{figure}[t!]
    \centering
    \begin{subfigure}{\linewidth}
        \centering
        \begin{tikzpicture}
        \begin{axis}[
            width=0.83\linewidth,
            xlabel = {Transmit Power $P$ (dBm)},
            ylabel = {Secrecy Rate (bps/Hz)},
            ymin = 0,
            ymax = 11,
            xmin = -10,
            xmax = 20,
            xtick = {-10,-5,0,5,10,15,20},         
            scaled y ticks=false, 
            grid = both,
            minor grid style={gray!25},
            major grid style={gray!50},
            legend columns=1, 
		legend entries ={$N=1$ PAS with AN, $N=1$ CAS with AN, $N=2$ PAS with AN, $N=2$ PAS without AN, $N=2$ CAS with AN},
            legend cell align = {left},
            legend style={font=\tiny},
            legend style={at={(0,1)},anchor=north west},
            legend image post style={scale=0.7}, 
            ]
            \addplot[
            blue,
            mark = triangle,
            mark repeat = 1,
            mark size = 2,
            line width = 1pt,
            style = solid,
            ]
            table[x index=0, y index=1, col sep=space]{Figures/single_pin_conv_SR.dat};
            \addplot[
            red,
            mark = triangle,
            mark repeat = 1,
            mark size = 2,
            line width = 1pt,
            style = solid,
            ]
            table[x index=0, y index=2, col sep=space]{Figures/single_pin_conv_SR.dat};
            \addplot[
            blue,
            mark = o,
            mark repeat = 1,
            mark size = 2,
            line width = 1pt,
            style = solid,
            ]
            table[x index=0, y index=1, col sep=space]{Figures/SR_pin_N_2_4_6_D_30.dat};
            \addplot[
            black,
            mark = o,
            mark repeat = 1,
            mark size = 2,
            line width = 1pt,
            style = solid,
            ]
            table[x index=0, y index=1, col sep=space]{Figures/SR_pin_N_2_4_D_30_noAN.dat};
            \addplot[
            red,
            mark = o,
            mark repeat = 1,
            mark size = 2,
            line width = 1pt,
            style = solid,
            ]
            table[x index=0, y index=1, col sep=space]{Figures/SR_conv_N_2.dat};
        \end{axis}
    \end{tikzpicture}
        \caption{}
        \label{fig:single_SR}
    \end{subfigure}
    
    \begin{subfigure}{\linewidth}
        \centering
        \begin{tikzpicture}
        \begin{axis}[
            width=0.83\linewidth,
            xlabel = {Transmit Power $P$ (dBm)},
            ylabel = {Secrecy Rate (bps/Hz)},
            ymin = 0,
            ymax = 15.5,
            xmin = -10,
            xmax = 20,
            xtick = {-10,-5,0,5,10,15,20},         
            scaled y ticks=false, 
            grid = both,
            minor grid style={gray!25},
            major grid style={gray!50},
            legend columns=1, 
		legend entries ={$N=4$ PAS with AN, $N=4$ PAS without AN, $N=4$ CAS with AN, $N=6$ PAS with AN, $N=6$ PAS without AN, $N=6$ CAS with AN},
            legend cell align = {left},
            legend style={font=\tiny},
            legend style={at={(0,1)},anchor=north west},
            legend image post style={scale=0.7}, 
            ]
            \addplot[
            blue,
            mark = triangle,
            mark repeat = 1,
            mark size = 2,
            line width = 1pt,
            style = solid,
            ]
            table[x index=0, y index=2, col sep=space]{Figures/SR_pin_N_2_4_6_D_30.dat};
            \addplot[
            black,
            mark = triangle,
            mark repeat = 1,
            mark size = 2,
            line width = 1pt,
            style = solid,
            ]
            table[x index=0, y index=2, col sep=space]{Figures/SR_pin_N_2_4_D_30_noAN.dat}; 
            \addplot[
            red,
            mark = triangle,
            mark repeat = 1,
            mark size = 2,
            line width = 1pt,
            style = solid,
            ]
            table[x index=0, y index=1, col sep=space]{Figures/SR_N_4_conv.dat};
            \addplot[
            blue,
            mark = o,
            mark repeat = 1,
            mark size = 2,
            line width = 1pt,
            style = solid,
            ]
            table[x index=0, y index=3, col sep=space]{Figures/SR_pin_N_2_4_6_D_30.dat};
            \addplot[
            black,
            mark = o,
            mark repeat = 1,
            mark size = 2,
            line width = 1pt,
            style = solid,
            ]
            table[x index=0, y index=3, col sep=space]{Figures/SR_pin_N_2_4_D_30_noAN.dat};
            \addplot[
            red,
            mark = o,
            mark repeat = 1,
            mark size = 2,
            line width = 1pt,
            style = solid,
            ]
            table[x index=0, y index=1, col sep=space]{Figures/SR_N_6_conv.dat};
        \end{axis}
    \end{tikzpicture}
        \caption{}
        \label{fig:multi_SR}
    \end{subfigure}
    \caption{Mean Secrecy Rate vs Transmit Power for various values of $N$.}
    \label{fig:figure2}
\end{figure}

\begin{figure}[t!]
    \centering
    \begin{subfigure}{\linewidth}
        \centering
        \begin{tikzpicture}
        \begin{axis}[
            width=0.83\linewidth,
            ylabel = {CDF},
            xlabel = {Secrecy Rate (bps/Hz)},
            ymin = 0,
            ymax = 1,
            xmin = 0,
            xmax = 15,
            xtick = {3,6,9,12,15},         
            scaled y ticks=false, 
            grid = both,
            minor grid style={gray!25},
            major grid style={gray!50},
            legend columns=1, 
		legend entries ={$N=1$ PAS with AN, $N=1$ CAS with AN, $N=2$ PAS with AN,$N=2$ PAS without AN, $N=2$ CAS with AN},
            legend cell align = {left},
            legend style={font=\tiny},
            legend style={at={(1,0)},anchor=south east},
            legend image post style={scale=0.7}, 
            ]
            \addplot[
            blue,
            mark size = 2,
            line width = 1pt,
            style = dashdotted,
            ]
            table[x index=0, y index=1, col sep=space]{Figures/single_pin_CDF.dat};
            \addplot[
            red,
            mark size = 2,
            line width = 1pt,
            style = dashdotted,
            ]
            table[x index=0, y index=1, col sep=space]{Figures/single_conv_CDF.dat};
            \addplot[
            blue,
            mark repeat = 1,
            mark size = 2,
            line width = 1pt,
            style = solid,
            ]
            table[x index=0, y index=1, col sep=space]{Figures/pin_2_CDF_rn.dat};
            \addplot[
            black,
            mark repeat = 1,
            mark size = 2,
            line width = 1pt,
            style = solid,
            ]
            table[x index=0, y index=1, col sep=space]{Figures/pin_2_CDF_no_rn.dat};
            \addplot[
            red,
            mark repeat = 1,
            mark size = 2,
            line width = 1pt,
            style = solid,
            ]
            table[x index=0, y index=1, col sep=space]{Figures/conv_2_CDF_rn.dat};
        \end{axis}
    \end{tikzpicture}
        \caption{}
        \label{fig:single_CDF}
    \end{subfigure}
    
    \begin{subfigure}
     {\linewidth}
        \centering
            \begin{tikzpicture}
        \begin{axis}[
            width=0.83\linewidth,
            ylabel = {CDF},
            xlabel = {Secrecy Rate (bps/Hz)},
            ymin = 0,
            ymax = 1,
            xmin = 0,
            xmax = 12,
            scaled y ticks=false, 
            grid = both,
            minor grid style={gray!25},
            major grid style={gray!50},
            legend columns=1, 
		legend entries ={$N=4$ PAS with AN, $N=4$ PAS without AN, $N=4$ CAS with AN, $N=6$ PAS with AN, $N=6$ PAS without AN, $N=6$ CAS with AN},
            legend cell align = {left},
            legend style={font=\tiny},
            legend style={at={(0,1)},anchor=north west},
            legend image post style={scale=0.5}, 
            ]
            \addplot[
            blue,
            mark repeat = 1,
            mark size = 2,
            line width = 1pt,
            style = dashdotted,
            ]
            table[x index=0, y index=1, col sep=space]{Figures/pin_4_CDF_rn.dat};
            \addplot[
            black,
            mark repeat = 1,
            mark size = 2,
            line width = 1pt,
            style = dashdotted,
            ]
            table[x index=0, y index=1, col sep=space]{Figures/pin_4_CDF_no_rn.dat};
            \addplot[
            red,
            mark repeat = 1,
            mark size = 2,
            line width = 1pt,
            style = dashdotted,
            ]
            table[x index=0, y index=1, col sep=space]{Figures/conv_4_CDF_rn.dat}; 
            \addplot[
            blue,
            mark repeat = 1,
            mark size = 2,
            line width = 1pt,
            style = solid,
            ]
            table[x index=0, y index=1, col sep=space]{Figures/pin_6_CDF_rn.dat}; 
            \addplot[
            black,
            mark repeat = 1,
            mark size = 2,
            line width = 1pt,
            style = solid,
            ]
            table[x index=0, y index=1, col sep=space]{Figures/pin_6_CDF_no_rn.dat}; 
            \addplot[
            red,
            mark repeat = 1,
            mark size = 2,
            line width = 1pt,
            style = solid,
            ]
            table[x index=0, y index=1, col sep=space]{Figures/conv_6_CDF_rn.dat}; 
            
        \end{axis}
    \end{tikzpicture}
        \caption{}
        \label{fig:multi_CDF}
    \end{subfigure}
    \caption{CDF of Secrecy Rates for various values of $N$, at $P=10$ dBm.}
    \label{fig:figure3}
\end{figure}
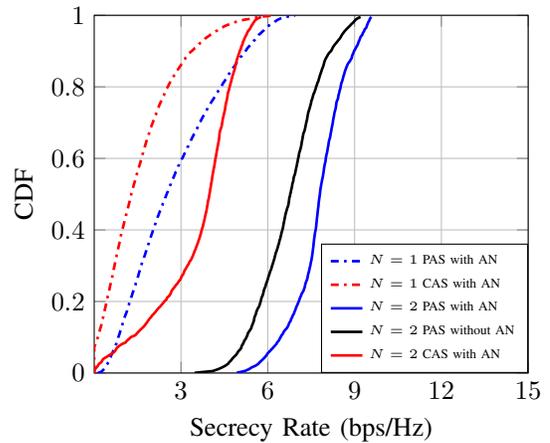
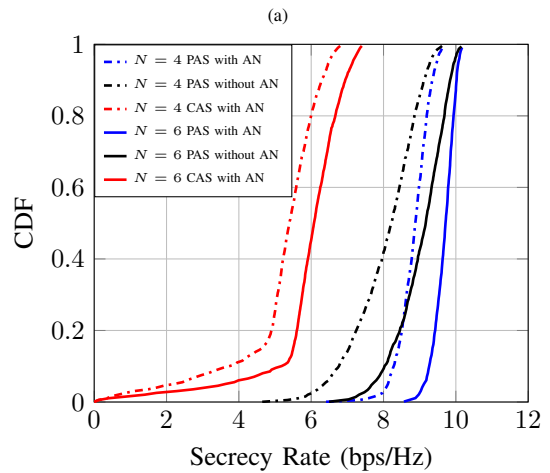

In this section, we numerically evaluate the performance of the proposed scheme by examining the mean achievable SR and the cumulative distribution function (CDF) of the SRs through Monte Carlo simulations. As benchmarks for comparison, we consider a PAS without AN \cite{huawai}, as well as a CAS, which is a conventional uniform linear array centered on and distributed along the $x$-axis. The simulation parameters are set as follows all dielectric waveguides are placed at a height of \( d = 3 \,\mathrm{m} \), with lengths equal to the side dimension \( D = 30 \,\mathrm{m} \), the noise power at both Bob and Eve is set to \( \sigma_B^2 = \sigma_E^2 = -90 \) dBm, the carrier frequency is set to \( f_c = 28\,\mathrm{GHz} \), and the effective refractive index of each dielectric waveguide is set to \( n_{\mathrm{eff}} = 1.4 \).

In Fig.~\ref{fig:figure2}, the mean SR is plotted as a function of transmit power for various values of \(N\) for both our proposed scheme and the benchmark approaches. More specifically, Fig.~\ref{fig:single_SR} focuses on the cases \(N=1\) and \(N=2\). Although \(N=1\) yields a notable gain over a CAS, the achievable SR remains limited in large indoor environments, where even with the reconfigurability of a PA, the channel gain difference between Bob and Eve can be minimal. In contrast, a significant SR increase is observed for \(N=2\), as the system can exploit both beamforming and AN capabilities that are not feasible for \(N=1\), where the parameters \(w\) and \(R_m\) are only scalars. In addition, the SR improvement for \(N=2\) continues at higher transmit power levels, significantly widening the performance gap with CAS. Finally, compared to a PAS that does not employ AN, the inclusion of AN provides a critical performance boost at higher transmit power levels, underscoring its key role in mitigating eavesdropping and ensuring robust secrecy performance. Moving to Fig.~\ref{fig:multi_SR}, it can be observed that the PAS with AN scheme consistently outperforms the corresponding CAS schemes, illustrating the effectiveness of the reconfigurability of PAs. Furthermore, when comparing PAS with and without AN, it is clear that the AN-enabled approach provides a significant performance gain over the entire transmit power range. In particular, our scheme with \(N=4\) even outperforms \(N=6\) PAS without AN when \(P\) exceeds 10 dBm. This result is particularly important because it demonstrates that higher performance can be achieved with fewer antennas by leveraging AN, thereby highlighting the potential for more cost-effective and energy-efficient secure communication designs.

Fig.~\ref{fig:figure3} shows the CDF of the SR for \(P = 10\) dBm, comparing the proposed and benchmark schemes over different values of \(N\). In particular, Fig.~\ref{fig:single_CDF} highlights the scenarios with \(N=1\) and \(N=2\). For \(N=1\), the proposed approach achieves not only a higher overall performance in terms of both median and peak SRs, but also an improved reliability. In particular, the CAS has a probability of more than 10\% of complete information leakage, i.e., \(\mathrm{SR} = 0\). Although the CAS with \(N=2\) achieves higher SRs, it still has a non-zero probability of zero SR, whereas the PAS allows Bob to maintain a non-zero SR even under unfavorable channel conditions. Moreover, incorporating AN into the PAS consistently outperforms the PAS benchmark throughout the CDF, especially in the lower SR region, highlighting the critical role of AN in enhancing system security and reliability. Moving on to Fig.~\ref{fig:multi_CDF}, we note that CAS still exhibits a non-negligible probability of \(\mathrm{SR}=0\), while both PAS schemes provide higher and more stable performance over various channel conditions. Nevertheless, the additional DoF provided by AN is critical to further improving the overall security performance in any configuration. In particular, it provides the greatest gains in the lower SR region of the CDF, thereby securing significant advantages under stronger Eve channels. This result underscores the importance of integrating AN into the PAS to maintain robust secrecy performance even in challenging eavesdropping environments.

\section{Conclusion}\label{sec:Conc}

In this paper, we introduced an AN–based beamforming scheme for secure downlink transmissions in PASs. Our approach jointly optimizes the beamforming vector, AN covariance, and the positions of the PAs to maximize the SR. A closed-form solution was derived for the single waveguide scenario, while an alternating optimization framework was developed for multi-waveguide configurations. Numerical results demonstrate that our scheme outperforms both the CAS and SotA PAS for PLS, highlighting the importance of integrating AN into PASs for enhanced security in next-generation wireless networks. These findings pave the way for future research in more complex environments with multiple Bobs and Eves.
 
\appendices

\section*{Appendix A - Proof of Lemma \ref{l1}}\label{ap:ferrari}
Since the objective function in \eqref{eq:opt_single_x} depends only on the scalar variable $\tilde{x}_{P}$, we differentiate it with respect to $\tilde{x}_{P}$ and set the derivative equal to zero to identify possible extrema. After some algebraic manipulation, the resulting first-order derivative is expressed as
\begin{equation}\label{eq:derivative_first}
\begin{aligned}
\frac{d\,\mathrm{SR}}{d\tilde{x}_{P}} =& \frac{\tilde{x}_{P}-x_b}{(\tilde{x}_{P}^2-2x_b \tilde{x}_{P}+K_2)^2+K_1(\tilde{x}_{P}^2-2x_b\tilde{x}_{P}+K_2)} \\
-&\frac{\tilde{x}_{P}-x_e}{(\tilde{x}_{P}^2-2x_e \tilde{x}_{P}+K_3)^2+K_1(\tilde{x}_{P}^2-2x_e\tilde{x}_{P}+K_3)},
\end{aligned}
\end{equation}
where $K_1=\lVert w\rVert^2 \eta/\sigma_B^2$, $K_2=x_b^2+y_b^2+d^2+\eta R_m/\sigma_B^2$ and $K_3=x_e^2+y_e^2+d^2+\eta R_m/\sigma_E^2$.
By setting \eqref{eq:derivative_first} equal to zero and performing some algebraic manipulations, we obtain the following quartic equation:
\begin{equation}\label{eq:derivative_pin}
\begin{aligned}
&\frac{d\,\mathrm{SR}}{d\tilde{x}_{P}} = 0 \;\Rightarrow \alpha_4\,\tilde{x}_{P}^4+\alpha_3\,\tilde{x}_{P}^3+\alpha_2\,\tilde{x}_{P}^2+\alpha_1\,\tilde{x}_{P}-\alpha_0=0, 
\end{aligned}
\end{equation}
with the coefficients of $\tilde{x}_{P}$ defined as 
\begin{align}
\alpha_4\,&=\,3 x_e - 3 x_b\nonumber,\\
\alpha_3\,&=\,4\,x_b^2 \;-\; 4x_e^2 \;+\; 2K_2 \;-\; 2K_3\nonumber,\\
\alpha_2\;&=\;K_1 x_e \;-\; K_1 x_b \;-\; 4K_2 x_b \;+\; 2K_3 x_b \nonumber\\
\;&-\; 2K_2 x_e \, +\; 4K_3 x_e \;+\; 4x_e^2 x_b \;-\; 4x_b^2 x_e,\\
\alpha_1\;&=\;K_2^2 \;-\; K_3^2 \;+\; K_1K_2 \;-\; K_1K_3 \nonumber\\
\;&+\; 4K_2x_bx_e \;-\; 4K_3x_bx_e\nonumber,\\
\alpha_0\;&=\; K_2^2 x_e \;+\; K_3^2 x_b \;+\; K_1K_3 x_b \;-\; K_1K_2 x_e.\nonumber
\end{align}
To simplify \eqref{eq:derivative_pin}, we perform the substitution  $\tilde{x}_{P}=\tilde{u}_{P}-\alpha_3/(4\alpha_4)$, which effectively eliminates the cubic term. Consequently, the equation can be rewritten in its depressed form as
\begin{equation}\label{eq:depressed}
    \tilde{u}_{P}^4+\alpha_2'\,\tilde{u}_{P}^2+\alpha_1'\,\tilde{u}_{P}+\alpha_0'=0,
\end{equation}
where the coefficients of $\tilde{u}_{P}$ are given as
\begin{align}
&\alpha_2'\,=\,\frac{\alpha_2}{\alpha_4}-\frac{3}{8}{\left(\frac{\alpha_3}{\alpha_4}\right)}^2,\\
&\alpha_1'\,=\,\frac{\alpha_1}{\alpha_4}-\frac{\alpha_2\alpha_3}{2\alpha_4^2}+\frac{1}{8}{\left(\frac{\alpha_3}{\alpha_4}\right)}^3,\\
&\alpha_0'\;=\;\frac{\alpha_0}{\alpha_4}+\frac{\alpha_3^2\alpha_2}{16\alpha_4^3}-\frac{\alpha_3\alpha_1}{4\alpha_4^2}-\frac{3}{256}{\left(\frac{\alpha_3}{\alpha_4}\right)}^4.
\end{align}
To solve \eqref{eq:depressed}, we need to factorize the quartic polynomial as the product of two quadratic terms, i.e.,
\begin{equation}\label{eq:factor}
    \tilde{u}_{P}^4+\alpha_2'\tilde{u}_{P}^2+\alpha_1'\tilde{u}_{P}+\alpha_0'=(\tilde{u}_{P}^2+p_1\tilde{u}_{P}+p_0)\,(\tilde{u}_{P}^2+q_1\tilde{u}_{P}+q_0).
\end{equation}
From the factorization in~\eqref{eq:factor}, matching the coefficients of each power of $\tilde{u}_P$ on both sides yields the following system of equations:
\begin{subequations}
\begin{align}
-p_1 = q_1&,\\
p_0q_0 = \alpha_0'&,\label{eq:pq}\\
p_1(q_0-p_0) = \alpha_1'&,\\
p_0 + q_0 - p_1^2 = \alpha_2'&.
\end{align}
\end{subequations}
Since solving this system directly is challenging, we introduce the substitution $\omega = q_0 - p_0.$ With this substitution, the system can be reformulated as 
\begin{subequations}
\begin{align}
p_1 =& -q_1 = \frac{\alpha_1'}{\omega}, \\
p_0 =& \frac{1}{2}\left(\alpha_2' - \omega + \frac{\alpha_1'^2}{\omega^2}\right), \label{eq:p0}\\
q_0 =& \omega + \frac{1}{2}\left(\alpha_2' - \omega + \frac{\alpha_1'^2}{\omega^2}\right). \label{eq:q0}
\end{align}
\end{subequations}
Substituting \eqref{eq:p0} and \eqref{eq:q0} into the \eqref{eq:pq} gives an equation for $m$:
\begin{equation}
    \omega^6+\beta_2\omega^4+\beta_1 \omega^2+\beta_0=0,
\end{equation}
where $\beta_2=4\,\alpha_0'-\alpha_2'^2$, $\beta_1=-2\alpha_2' \alpha_1'^2$ and $\beta_0=-\alpha_1'^4$. By setting $l = \omega^2$, the equation simplifies to the following cubic form:
\begin{equation}
    l^3+\beta_2l^2+\beta_1l+\beta_0=0.
\end{equation}
To eliminate the quadratic term, we introduce the substitution $l=z-\beta_2/3$, which transforms the polynomial into
\begin{equation}\label{eq:cardano}
    z^3+\beta_1'z+\beta_0'=0,
\end{equation}
where $\beta_1'=\beta_1+\beta_2^2/3$ and $\beta_0'=2/(3\beta_2)^3-\beta_1\beta_2/3+\beta_0$. Equation~\eqref{eq:cardano} is now in the depressed cubic form and can be solved using Cardano's formula. In this formulation, the discriminant is defined as $\Delta=\left(\beta_0'/2\right)^2+\left(\beta_1'/3\right)^3$, which indicates the nature of the roots. The real solutions for $z$ are given by
\begin{equation}
z=\left\{
\begin{aligned}
&\sqrt[3]{-\frac{\beta_0'}{2}+\sqrt{\Delta}} + \sqrt[3]{-\frac{\beta_0'}{2}-\sqrt{\Delta}},\quad \Delta\geq0,\\
&2\sqrt{-\frac{\beta_1'}{3}}\cos{\left(\frac{\theta+2\pi k}{3}\right)}, \, k=0,1,2,\quad \Delta<0,
\end{aligned}
\right.
\end{equation}
where 
\begin{equation}
    \theta=\arccos{\left(\frac{-\frac{\beta_0'}{2}}{\sqrt{-\left(\frac{\beta_1'}{3}\right)^3}}\right)}.
\end{equation}
It should be noted that although the case $\Delta < 0$ yields three distinct real roots, for our purposes a single real solution for $z$ is sufficient to facilitate the factorization of $\tilde{u}_{P}$. Once $z$ is determined, we reverse the substitutions to recover $l$ and then $\omega$, with only one value of $\omega$ required to complete the factorization in equation~\eqref{eq:factor}. Next, we solve the two resulting quadratic equations from the second part of \eqref{eq:factor} to obtain their roots and substitute these values to determine the candidate values of $\tilde{x}_{P}$. Finally, we select the value that maximizes our objective function while satisfying the waveguide boundary constraint $\mathrm{C}_1$, which completes the proof.

\section*{Appendix B - Proof of Lemma \ref{l2}}\label{ap:monotonic}

The function in \eqref{eq:SR_fixed_x} depends on a single variable, $R_m$. To evaluate its monotonicity, we differentiate the function for $R_m$. After algebraic manipulations, we obtain the derivative as
\begin{equation}
    \frac{d\,\mathrm{SR}}{dR_m} =  \frac{\eta\,\left(r_B^2\sigma_B^2 - r_E^2\sigma_E^2\right)}{(\eta R_m + r_B^2\sigma_B^2)(\eta R_m + r_E^2\sigma_E^2)}.
\end{equation}
Since $\eta$ and the denominator are strictly positive, the sign of the derivative depends only on the term $r_B^2\sigma_B^2 - r_E^2\sigma_E^2$, which is independent of $R_m$. Consequently, the derivative keeps a constant sign, and the monotonicity of $\mathrm{SR}$ is characterized by
\begin{equation}
\left\{
\begin{aligned}
r_B^2\sigma_B^2 > r_E^2\sigma_E^2 &\quad \Longrightarrow \quad \mathrm{SR}\uparrow, \\
r_B^2\sigma_B^2 < r_E^2\sigma_E^2 &\quad \Longrightarrow \quad \mathrm{SR}\downarrow.
\end{aligned}
\right.
\end{equation}
Since the derivative has a constant sign, the function has no stationary points. Thus, any extremum must occur at one of the endpoints of $R_m$, namely at $0$ and $P$, which completes the proof.

\bibliographystyle{IEEEtran}
\bibliography{references}

\end{document}